\documentclass[12pt]{article}

\global\arraycolsep=1pt
\usepackage{amssymb}
\usepackage{amsmath}
\usepackage{amscd}
\usepackage {pstricks}
\newcommand{\beq}{\begin{equation}}
\newcommand{\eeq}{\end{equation}}
\newcommand{\beqa}{\begin{eqnarray}}
\newcommand{\eeqa}{\end{eqnarray}}
\newcommand{\CR}{\nonumber \\}

\newcommand{\bq}{\mathbf{q}}
\newcommand{\qbinom}[2]{\genfrac{[}{]}{0pt}{}{#1}{#2}_\bq}

\newcommand{\Exp}[1]{\exp\left\{#1\right\}}

\newcommand{\proof}{\noindent{\it Proof.\hskip10pt}} 
\newcommand{\qed}{\hfill\fbox{}}
\newcommand{\Setv}[2]{ \left\{\, #1\, \vert\, #2\, \right\} }
\newcommand{\bN}{{\mathbb N}}
\newcommand{\bZ}{{\mathbb Z}}
\newcommand{\qint}[2]{[\, #1\, ]_{#2}}
\newcommand{\qintf}[2]{[\, #1\, ]!_{#2}}
\newcommand{\qbin}[3]{\left[{#1\atop #2}\right]_{#3}}
\newcommand{\be}{\begin{equation}}
\newcommand{\ee}{\end{equation}}
\newcommand{\ba}{\begin{eqnarray}}
\newcommand{\ea}{\end{eqnarray}}
\renewcommand{\theequation}{\thesection.\arabic{equation}}
\newcommand{\Section}{\setcounter{equation}{0} \section}
\renewcommand{\thefootnote}{\fnsymbol{footnote}}
\begin{document}
%
\begin{titlepage}
\begin{flushright}
{April, 2009}
\end{flushright}
\vspace{0.5cm}
\begin{center}
{\Large \bf
Quiver Matrix Model and Topological \\
 Partition Function in Six Dimensions}
\vskip1.0cm
{\large Hidetoshi Awata and Hiroaki Kanno}
\vskip 1.0em
{\it 
Graduate School of Mathematics \\
Nagoya University, Nagoya, 464-8602, Japan}
\end{center}
\vskip1.5cm

\begin{abstract}
We consider a topological quiver matrix model which is
expected to give a dual description of the instanton dynamics 
of topological $U(N)$ gauge theory on $D6$ branes. 
The model is a higher dimensional analogue of the ADHM matrix model
that leads to Nekrasov's partition function. 
The fixed points of the toric action
on the moduli space are labeled by colored plane partitions. 
Assuming the localization theorem, we compute 
the partition function as an equivariant index. 
It turns out that the partition function does not depend on 
the vacuum expectation values of Higgs fields that break
$U(N)$ symmetry to $U(1)^N$ at low energy. 
We conjecture a general formula of the partition function,
which reduces to a power of  the MacMahon function, 
if we impose the Calabi-Yau condition.
For non Calabi-Yau case we prove the conjecture 
up to the third order in the instanton expansion. 
\end{abstract}
\end{titlepage}


\renewcommand{\thefootnote}{\arabic{footnote}} \setcounter{footnote}{0}


\Section{Introduction}

During the recent developments in
the non-perturbative dynamics of supersymmetric gauge/string theories,
we have witnessed many examples of the topological partition function which are exactly computable. 
They arise from the enumerative problems and are defined as the generating functions of 
instanton or BPS state counting.  Thus they carry useful information
for testing various dualities in supersymmetric theories,
such as mirror symmetry, electro-magnetic duality and gauge/string correspondence.
One of the important mathematical ideas in these computations is 
the equivariant localization theorem and it has revealed a close relation 
to the combinatorics. We use basic combinatorial tools in representation theory, 
such as the partition (the Young diagram),
the plane partition, the Schur function and the Macdonald function.
For example, $N$-tuple of the Young diagrams or the colored partition appears 
in Nekrasov's computation of Seiberg-Witten prepotential \cite{Nek, NO,NY1}. 
We use the (skew) Schur function to write down the topological
vertex  \cite{AMV, AKMV}, which gives a building block of topological
string amplitudes on toric Calabi-Yau threefolds. 
It is also related to the plane partition \cite{ORV}. 
The generating function of counting plane partitions is the MacMahon function,
which is ubiquitous in topological gauge/string theory. For example, 
it appears in topological string amplitude on the conifold \cite{GV}, 
the Gopakumar-Vafa invariants \cite{GVM}
and the Donaldson-Thomas theory \cite{DT, INOV, MNOP}. 
Finally the Macdonald function, which is the most general class of the
symmetric functions, was employed to construct
a refinement of the topological vertex \cite{AK1, IKV, Taki, AK2}.

It is quite interesting that the topological  partition function often takes
the plethystic form\footnote{The plethystic exponential also appears
in the problem of counting gauge invariant operators in quiver gauge theories.
See for example \cite{BFHH}.}. 
Namely there exists a function ${\cal F}(t_1, t_2, \cdots)$ and the partition function is
written as the plethystic exponential;
\beq
Z_{\mathrm{top}}(t_1, t_2, \cdots) = \exp 
\left(
\sum_{k=1}^\infty \frac{1}{k} {\cal F}(t_1^k, t_2^k, \cdots)
\right)~,
\eeq
where we have denoted parameters of the theory collectively  as $(t_1, t_2, \cdots)$.
This implies that the partition function has an infinite product (Euler product) form.
Topological string amplitudes in the Gopakumar-Vafa form are basic examples.
Furthermore the fact that Nekrasov's partition function allows an expansion
in the plethystic form is crucial to identify it 
as topological string amplitudes or their refined version \cite{AK1, AK2}.
The MacMahon function which is a basic partition
function in the Donaldson-Thomas theory has also
the plethystic form;
\beq
M(t) := \prod_{n=1}^\infty (1-t^n)^{-n} = \exp
\left(
\sum_{k=1}^\infty \frac{1}{k} \frac{1}{(t^{\frac{k}{2}} - t^{-\frac{k}{2}})^2}
\right)~.
\eeq
It is an interesting challenge to uncover a possible mathematical and/or physical origin 
of the plethystic exponential in general.

In this paper we consider a topological quiver matrix model
which is expected to describe low energy and instanton dynamics of
the topological gauge theory on $D6$ branes \cite{Jaf, CSS}. 
The model is a six dimensional analogue of the ADHM matrix model 
derived from low energy effective theory of $D4$-$D0$ system
\cite{WitADHM1,WitADHM2,Doug1,Doug2}. 
The instanton partition function for the ADHM matrix model is nothing but Nekrasov's
partition function which is related to the Seiberg-Witten
prepotential and topological string. 
The fixed points of the toric action on the moduli space of 
the ADHM matrix model are labeled by colored partitions \cite{Nak}.
Hence the localization theorem tells us that the partition function can be 
computed as a summation over colored partitions.
We consider a similar instanton partition function
for the six dimensional quiver matrix model.
To compute the partition function based on the localization
theorem, which we assume throughout the paper,
we introduce $T^3$ action $(z_1, z_2, z_3) \to (e^{i\epsilon_1} z_1, 
e^{i\epsilon_2} z_2, e^{i\epsilon_3} z_3)$ on ${\mathbb C}^3$,
which may be regarded as the $\Omega$ background of Nekrasov. 
We also consider the action of the maximal torus
$U(1)^N$ of the gauge group $U(N)$, where the rank $N$ refers 
to the number of $D6$ branes.
These toric actions induce the action on the moduli space of the topological
quiver matrix model and the fixed points are labeled by $N$-tuples of 
plane partitions (3d Young diagrams), or colored plane partitions \cite{CSS}. 
The partition function is defined as an equivariant index and
thus a rational function of equivariant parameters $q_i = e^{i\epsilon_i}$ from
$T^3$ action and $e_\alpha = e^{ia_\alpha}$ from the maximal torus.
Physically $a_\alpha$ are the vacuum expectation values of Higgs fields. 
We call the condition $\bq:= \sqrt{q_1 q_2 q_3}=1$ Calabi-Yau condition,
which can be compared with the anti-self-duality in four dimensions.
When the Calabi-Yau condition is imposed,
the weight or the measure at any fixed point is $\pm 1$ and
consequently the partition function of $U(N)$ theory 
reduces to the $N$-th power of the MacMahon function. 
Hence the partition function is independent of both $q_i$ and 
$e_\alpha$.

In non Calabi-Yau case the weight at each fixed point becomes
rather complicated expression and the partition function does depend on
the equivariant parameters $q_i$. However, we have found that even in this case,  
the instanton partition function is still independent of  $e_\alpha$
for lower instanton numbers. We believe this is quite surprising. 
Based on explicit computations for lower rank $N$ and instanton number $k$, 
we propose the following formula of the topological partition function;
\beq
Z_{\mathrm{6D}}^{U(N)} (q_i; \Lambda) = \exp 
\left(
\sum_{n=1}^\infty \frac{1}{n} F_N (q_1^n, q_2^n, q_3^n; \Lambda^n)
\right)~, \label{pf}
\eeq
where
\beq
F_N := - \frac{\widetilde\Lambda(1+ \bq^2+ \bq^4 + \cdots + \bq^{2N-2})}
{(1-\widetilde\Lambda)(1- \bq ^{2N} \widetilde\Lambda)}
\frac{(1-q_1 q_2)(1- q_2 q_3)(1- q_3 q_1)}{(1-q_1) (1-q_2) (1-q_3)}~, \label{fe}
\eeq
and $\widetilde\Lambda:= (-\bq)^{-N} \Lambda$ is a renormalized parameter 
of the instanton expansion parameter $\Lambda$.
Note that when $\bq=1$, $q_i$ dependence of the partition function disappears completely 
and we have $Z_{CY3}^{U(N)}=M(\widetilde\Lambda)^N$. 
If we expand the first factor of $F_N$ in $\widetilde\Lambda$, the coefficients are $\bq$-integers.
Thus the above formula may have a certain interpretation of $\bq$-deformation.

Mathematically the topological partition function we compute in this paper 
has a natural meaning in $K$ theory. 
The $K$ theoretic version of Nekrasov's partition function is physically regarded as
a five dimensional lift and it is the $K$ theoretic version which we can identity with
topological string amplitudes. Thus we expect that our partition function has
a seven dimensional interpretation, if we combine it with the perturbative contributions
from Kaluza-Klein modes.
In fact the partition function \eqref{pf} for abelian theory $(N=1)$ was conjectured in \cite{TN} 
together with a curious relation to $M$ theory partition function. 
It is possible that the simplicity of the topological partition function we proposed above
originates from the maximally supersymmetry of Yang-Mills gauge theory. 
In fact we have encountered before a similar example 
in five dimensional $U(1)$ gauge theory with an adjoint hypermultiplet \cite{AK2, IKS, PS, AK3}. 
In this case Nekrasov's partition function takes the following form of the plethystic exponential;
\beqa
&&
Z_{\mathrm{5D}}^{\mathrm{adj}} (q_i, Q ; \Lambda) =
\sum_{\lambda}\Lambda^{|\lambda|} 
\prod_{s\in\lambda}
\frac
{1-Q q_1^{ -a(s)  } q_2^{ \ell(s)+1}}{{1-  q_1^{ -a(s)  }} q_2^{ \ell(s)+1}}
\frac
{1-Q q_1^{a(s)+1} q_2^{-\ell(s)  }} {{1-q_1^{a(s)+1}} q_2^{-\ell(s) }}\CR
&&~~=
\exp\left\{\sum_{n>0} 
\frac{1}{n}
\frac{\Lambda^n}{1- Q^n \Lambda^n}
\frac{(1- Q^n q_1^n)(1- Q^n q_2^{n})}
{(1- q_1^{n})(1- q_2^{n})}
\right\}. \label{5DU1}
\eeqa
The left hand side of \eqref{5DU1} is a summation over partitions $\lambda$.  
The integers $a(s)$ and $\ell(s)$ in the product are the arm length and 
the leg length that are defined by the corresponding Young diagram. 
The mass $m$ of the adjoint hypermultiplet defines the parameter $Q = e^{-m}$
of the mass deformation. It is tempting to compare the parameter $\bq$ 
in \eqref{fe} with $Q$ in \eqref{5DU1}.

The topological quiver matrix model in this paper is a
$0+1$-dimensional \lq\lq world-line\rq\rq\ theory on $D0$ branes.
However, we should emphasize that if the Donaldson-Thomas 
theory is formulated as a topological gauge theory on $D6$ branes, 
the effect of $D2$-branes on $D6$ cannot be negligible. 
If we want to regard the topological quiver matrix model as a dual description of
the Donaldson-Thomas theory, it has to accommodate $D2$ branes. 
For example the contribution of \lq\lq 0-2\rq\rq\ string should appear as a multiplicative
factor to the partition function \eqref{pf}. 
Recall that in \cite{ORV} the enumeration of plane partitions
led to the generating function $Z_{\lambda \mu \nu}(u)
= M(u) C_{\lambda \mu \nu}(u)$, where three partitions 
(Young diagrams) $\lambda, \mu, \nu$ define
asymptotic conditions on plane partitions and $u := e^{-g_s}$
is related to the string coupling $g_s$. 
The generating function is given by the topological vertex 
$C_{\lambda \mu \nu}(u)$ and the MacMahon function
appears as a normalization factor. 
From this viewpoint what we have proposed above is 
an extension of the MacMahon factor to non Calabi-Yau case.
Thus the issue is closely related to the problem of 
extending the topological vertex to toric K\"ahler threefolds.
In any case incorporating the effect of $D2$ branes to the quiver matrix
model is beyond the scope of the present paper. 
We want to address this issue in future.

The paper is organized as follows;
in section 2 we review the construction of topological quiver matrix model
following \cite{Jaf, CSS, MNS1, MNS2}.
We clarify the relation of the stability condition and the vanishing theorem,
which was not emphasized before in \cite{Jaf, CSS}. 
In section 3 we introduce the toric action on the moduli space of the 
topological matrix model. The fixed points are isolated and they are 
classified by $N$-tuples of plane partitions (colored plane partitions). 
Hence we can compute the partition function which is defined as an equivariant index
by summing up the contribution at each colored plane partition.
When the Calabi-Yau condition is imposed,  the partition function
reduces to a power of MacMahon function.
Section 4 is the main part of the paper.
We compute the partition function for non Calabi-Yau case
and prove our conjecture \eqref{pf} up to instanton number three.
We first look at possible poles of the partition function in $e_\alpha$
and show that all the residues vanish. Hence the partition function
is independent of $e_\alpha$ and we compute it 
by taking appropriate limit (the decoupling limit). 
It turns out that the identities among $\bq$-integers derived from 
the $\bq$ binomial theorem reduce
the conjecture for $U(N)$ theory to that for $U(1)$ theory.
We can check the conjecture for $U(1)$ theory by
direct computation. In Appendix A we give a brief review of the ADHM matrix model 
and the relation to Nekrasov's partition function.
In Appendix B we check the well-definedness of our partition function. 


\Section{Topological quiver matrix model}

Let us consider a quiver matrix model for topological gauge theory
on $D6$ brane \cite{Jaf,CSS}. This is an analogue of the ADHM matrix model.
As in the ADHM construction we introduce two vector spaces $V$ and $W$ of 
complex dimensions $\dim_{\mathbb C} V =k$ and $\dim_{\mathbb C} W = N$.
From the perspective of the gauge theory on the world volume of $D6$ brane,
$k$ is the number of $D0$ branes (the instanton number)
and $N$ is the number of $D6$ branes (the rank). 
The basic fields are the matrices
\beq
B_1, B_2, B_3, \varphi \in \mathrm{Hom}~(V,V)~, \qquad
I \in \mathrm{Hom}~(W,V)~,
\eeq
where $(B_1, B_2, B_3, \varphi)$ come from \lq\lq 0-0\rq\rq\ string and $I$ comes from \lq\lq 6-0\rq\rq\ string.
Compared with the ADHM date for four dimensional gauge theory,
the model does not have $J \in  \mathrm{Hom}~(V, W)$, 
or \lq\lq 0-6\rq\rq\ string\footnote{See \cite{OY} for an explanation from the viewpoint of $T$ duality.}. 
Instead, we have a matrix $\varphi  \in \mathrm{Hom}~(V,V)$. 
This additional field comes from a reduction of a topological theory in eight dimensions  \cite{BKS}, 
which originates from the ten dimensional super Yang-Mills theory. 
We can show the vanishing theorem which implies that $\varphi=0$ on the moduli space \cite{BKS}.
Hence, classically $\varphi$ is irrelevant to the moduli problem. However,
the presence of $\varphi$ is crucial for imposing the constraint \eqref{Bterm} 
to be introduced below.
We consider the following equations of motion\footnote{See also \cite{Wit2, Ohta, HIO} on 
the equations of ADHM type for $D0$-$D6$ and $D0$-$D8$ systems.
In these papers a background $B$-field was introduced to obtain BPS bound states. }
\beqa
{\cal E}_{F}&:=& [ B_i, B_j ] + \epsilon_{ijk} [ B_k^\dagger, \varphi] = 0, \label{Fterm}\\
{\cal E}_{D}(\zeta) &:=&\sum_{i=1}^3 [B_i, B_i^\dagger] 
+ [\varphi, \varphi^\dagger] + I I^\dagger - \zeta \cdot {\bf 1}_{k \times k}=0, \quad (\zeta> 0), \label{Dterm}\\
{\cal E}_{B}&:=& I^\dagger \varphi = 0~. \label{Bterm}
\eeqa
These are the gauge fixing conditions or the constraints
in our topological matrix model with gauge symmetry $U(k)$. 
Among them \eqref{Fterm} gives three $F$-tern (holomorphic) conditions and
\eqref{Dterm} is the (real) $D$-term condition which is responsible for the stability.
There is no counter part of \eqref{Bterm} in the ADHM equation and 
it may be interesting to clarify its implication.
Since $\varphi$ describes the normal direction to the 
world volume of $D6$ branes, \eqref{Bterm} means that the \lq\lq 6-0\rq\rq\ string $I$ is
orthogonal to the normal direction \cite{Jaf}. This implies that $D0$ branes are forced to 
be bound to $D6$ branes. The reason why we should impose the constraint \eqref{Bterm} 
might be related to the fact that $D6$-$D0$ system cannot make a BPS bound state 
without an appropriate flux along the $D6$ branes \cite{Wit2}. In any case
it is important to further clarify a possible explanation from the viewpoint of the BPS states.

One can construct a topological matrix model following the prescription 
of \cite{MNS1, MNS2}. This was achieved in \cite{Jaf, CSS}. 
Since we have the constraints \eqref{Fterm}-\eqref{Bterm}, 
the moduli space of the topological theory is identified with
\beq
{\cal M}_{\mathrm{TQM}} := \{ (B_i, \varphi, I)~\vert~
{\cal E}_{F} = {\cal E}_{D}(\zeta) = {\cal E}_{B} =0 \} / U(k)~. \label{symplectic}
\eeq
Let us count the degrees of freedom. Since $B_i, \varphi$ and $I$ are complex matrices,
there are $8 k^2 + 2Nk$ degrees of freedom. The constraints impose
$6 k^2 + k^2 + 2Nk$ relations. Finally we have $U(k)$ gauge symmetry.
Hence the formal dimension of the moduli space ${\cal M}_{\mathrm{TQM}}$ vanishes.
This is certainly due to the fact that the origin of our theory is the ten dimensional super Yang-Mills theory
which is maximally supersymmetric. 
However, this gives us a puzzle, since we naively expect $6Nk$ degrees of freedom
for $k$ $D0$ branes bound to $N$ $D6$ branes, which means the (complex) dimensions of
the moduli space are $3Nk$. We suspect the following fact is related to this issue. 
According to the general theory of topological matrix model
\cite{MNS1, MNS2}, we are computing the Euler character of 
the anti-ghost bundle on the moduli space ${\cal M}_{\mathrm{TQM}}$.
However, in the present case, 
the anti-ghost bundle only makes sense as a complex of bundles \cite{Jaf}
and it is in fact different from the tangent bundle of ${\cal M}_{\mathrm{TQM}}$
which is defined by the linearization of the constraints.

In the ADHM matrix model the moduli space of the type \eqref{symplectic} 
comes from the hyperK\"ahler quotient construction. It is well known that
we have an equivalent definition of the moduli space 
by affine algebro-geometric quotient \cite{NY1,Nak},
where we omit the $D$-term condition but impose the stability condition on the orbits. 
We also have to complexify the gauge group to $GL(k, {\mathbb C})$. 
Since it is not established at the moment that a similar equivalence holds 
in our higher dimensional generalization, 
we assume it and we take
\beq
\widetilde{\cal M}_{\mathrm{TQM}} := \{ (B_i, \varphi, I)~\vert~
{\cal E}_{F} = {\cal E}_{B} =0,~~\mathrm{stability~condition} \} 
/\!/ GL(k, {\mathbb C})~, \label{AGquotient}
\eeq
as our definition of the moduli space in the following.
In \eqref{AGquotient} $/\!/$ means the affine algebro-geometric quotient
where we only consider the orbits that satisfy the stability condition;
\beq
\mathrm{There~is~no~proper~subspace}~S \subsetneq  V \quad
\mathrm{with} \quad B_i S \subset S, \quad \mathrm{Im}(I) \subset S~.  \label{stability}
\eeq
We now show that under the stability condition \eqref{stability} we have
a vanishing theorem that $\varphi=0$, if $(B_i, \varphi, I) \in \widetilde{\cal M}_{\mathrm{TQM}}$.
Firstly, we note that the $F$-term condition ${\cal E}_{F}=0$ splits into two independent equations,
\beq
[ B_i, B_j ] = [B_k^\dagger, \varphi] = 0~. \label{split} 
\eeq
To see it, let $A_k := [B_k^\dagger, \varphi] \in \mathrm{Hom}~(V,V)$. 
Then by \eqref{Fterm} and the Jacobi identity, we have
\beq
\mathrm{Tr}~A_k^\dagger A_k 
= \frac{1}{2} \epsilon_{ijk} \mathrm{Tr}~[\varphi^\dagger, B_k] [B_i, B_j]
=  \frac{1}{2} \epsilon_{ijk} \mathrm{Tr}~\varphi^\dagger [ [B_i, B_j],  B_k] 
=0~.
\eeq
Hence $A_k=0$ and \eqref{split} holds. To prove the vanishing theorem
it is enough to show that $\varphi^\dagger v = 0$ for any $v \in V$. 
By the stability condition the vector space $V$ is generated by 
applying $B_i$'s on $\mathrm{Im}(I)$. Hence any 
vector $v \in V$ can be written as $v = B_{i_1} B_{i_2} \cdots B_{i_n} I(w)$
by choosing an appropriate vector $w \in W$. Since $\varphi^\dagger$ and $B_i$'s
commute by \eqref{split}, $\varphi^\dagger v 
= B_{i_1} B_{i_2} \cdots B_{i_n} \varphi^\dagger I(w) =0$,
where the last equality follows from ${\cal E}_{B} =0$. 
This completes the proof of the vanishing theorem. 
Consequently the matrix $\varphi$ decouples and
the moduli space is actually 
\beq
\widetilde{\cal M}_{\mathrm{TQM}} = \{ (B_i, I)~\vert~
[B_i, B_j] = 0,~~\mathrm{stability~condition} \} 
/\!/ GL(k, {\mathbb C})~. \label{finalmoduli}
\eeq
Note that the matrix $I$ only concerns the stability condition.

It follows from \eqref{finalmoduli} that when $N=1$, we can identify 
$\widetilde{\cal M}_{\mathrm{TQM}}$ with the Hilbert scheme
of $k$ points in ${\mathbb C}^3$;
\beq
\mathrm{Hilb}^k ({\mathbb C}^3) 
= \{ \mathcal{I} \subset {\mathbb C}[z_1, z_2, z_3]~\vert~
\dim \left(  {\mathbb C}[z_1, z_2, z_3] / \mathcal{I} \right) = k \}~,
\eeq
where $\mathcal{I}$ denotes an ideal in the polynomial ring
$ {\mathbb C}[z_1, z_2, z_3]$.
We note that ${\mathbb C}[B_1, B_2, B_3] \simeq 
 {\mathbb C}[z_1, z_2, z_3]$, since $[B_i, B_j] =0$. 
 For any ideal $\mathcal{I} \in 
\mathrm{Hilb}^k ({\mathbb C}^3)$, let
$V = {\mathbb C}[z_1, z_2, z_3] / \mathcal{I}$.
We define $B_i \in \mathrm{Hom}~(V,V)$ 
by the multiplication of $z_i$ modulo $\mathcal{I}$.
When $N=1$, $I$ is defined by giving an element $I(1) \in V$.
We take $I(1) = 1$ modulo $\mathcal{I}$. 
Then clearly $[B_i, B_j]=0$ and it is easy to see
that the stability condition is satisfied. Conversely,
for any element $(B_i, I) \in \widetilde{\cal M}_{\mathrm{TQM}}$,
we define a map $\mu: {\mathbb C}[z_1, z_2, z_3] \to V$
by $\mu(f(z_1, z_2, z_3)) := f(B_1, B_2, B_3)\cdot I(1)$,
which is well-defined thanks to $[B_i, B_j]=0$.
The stability condition implies that $\mu$ is surjective.
Hence, if we define an ideal in ${\mathbb C}[z_1, z_2, z_3]$ 
by $\mathcal{I} := \mathrm{Ker}~\mu$, then
${\mathbb C}[z_1, z_2, z_3] / \mathcal{I} \simeq V$. 
Since $\dim_{\mathbb C} V = k$, we have 
$\mathcal{I} \in \mathrm{Hilb}^k ({\mathbb C}^3)$.

We can write down the deformation complex associated with 
the moduli space \eqref{AGquotient} by the standard manner;
\beq
\begin{array}{ccccc}
 & & \oplus_{k=1}^3 \mathrm{Hom}~(V, V)_{k}  & & \\
 & & \oplus& & \oplus_{i,j=1}^3 \mathrm{Hom}~(V,V)_{[ij]} \\
\mathrm{Hom}~(V,V) &~~\overset{\sigma}{\longrightarrow} &
\mathrm{Hom}~(V,V) & \overset{\tau}{\longrightarrow}~~& \oplus \\ 
 & & \oplus& & \mathrm{Hom}~(V,W) \\
 & & \mathrm{Hom}~(W,V) & & 
\end{array},
\eeq
where the first term corresponds to the degrees of freedom of infinitesimal
gauge transformation, the middle term parametrizes the tangent space of $\widetilde{\cal M}_{\mathrm{TQM}}$
and the last term comes from the linearization of the constraints \eqref{Fterm} and \eqref{Bterm}. 
At a point $(B_i, \varphi, I) \in \widetilde{\cal M}_{\mathrm{TQM}}$
the maps $\sigma$ and $\tau$ are defined 
by 
\beqa
\sigma(\phi) &:=& \delta_\phi(B_i, \varphi, I)
= ([\phi, B_i], [\phi, \varphi], \phi I), \\
\tau((\delta B_i,\delta\varphi, \delta I)) &:=&
\left(  [\delta B_i, B_j] + [B_i, \delta B_j] + \epsilon_{ijk}
( [\delta B_k^\dagger, \varphi] + [B_k^\dagger, \delta\varphi] ),
\delta I^\dagger \varphi + I^\dagger \delta \varphi \right). \CR
&&
\eeqa
Note that the gauge invariance of the constraints implies $\tau \circ \sigma =0$. 


\Section{Instanton partition function}

Generalizing the computation of Nekrasov's partition function as the topological
partition function of the ADHM matrix model, which is reviewed in Appendix A, 
we want to compute the partition function of our quiver matrix model.
By introducing the toric action on the moduli space and applying the localization theorem
the partition function is computed as an equivariant index. 
We consider two toric actions on the moduli space $\widetilde{\cal M}_{\mathrm{TQM}}$.
The first one comes from the canonical $T^3$ action
$(z_1, z_2, z_3) \to (e^{i\epsilon_1} z_1, e^{i\epsilon_2} z_2, e^{i\epsilon_3} z_3)$
on ${\mathbb C}^3$, which is an example of the $\Omega$ background of Nekrasov.
The second one is induced from the action of the maximal torus $U(1)^N$ 
of the global gauge group $U(N)$. Physically the parameters $a_\alpha,(\alpha=1, \cdots, N)$ 
of the maximal torus correspond to the vacuum expectation values of the Higgs scalars
or the distances of $D6$ branes. In the following we
use the notations $q_i := e^{i\epsilon_i}, e_\alpha := e^{i a_\alpha}$. 
Since we have $GL(k, {\mathbb C})$ gauge symmetry, 
the condition of the fixed point is imposed up to gauge transformations.
Hence, the conditions we have to solve are
\beqa
q_j B_j &=& g(q_i, \lambda) \cdot B_j \cdot g^{-1}(q_i, \lambda)~, \qquad (j=1,2,3) \label{fixB} \\
q_1 q_2 q_3 \varphi &=& g(q_i, \lambda) \cdot \varphi \cdot g^{-1}(q_i, \lambda)~ , \label{fixphi} \\
I \cdot \lambda &=& g(q_i, \lambda)\cdot  I~,  \label{fixI} 
\eeqa
where $\lambda \in U(1)^N$. 
Note that at each fixed point the conditions \eqref{fixB}-\eqref{fixI} 
define a homomorphism $g : T^3 \times U(1)^N \to GL(k, {\mathbb C})$ . 
By the homomorphism $g$ we can regard the vector spaces $W$ and $V$,
which were originally $GL(k, {\mathbb C})$ modules, as $T^3 \times U(1)^N$ modules. 
Through the matrix $I$ the action of the maximal torus $U(1)^N$ on $W$ is
translated into a $U(1)^N$ action on $V$. It is helpful to keep these points in mind, 
when we compute the equivariant character of the deformation complex. 
In the following we will identify one dimensional $T^3 \times U(1)^N$ modules
with the equivariant parameters of the toric action. Namely $q_i$ and $e_\alpha$ stand for
the module where  $T^3 \times U(1)^N$ acts as the multiplication of $q_i$ and $e_\alpha$,
respectively. Hence a product of the equivariant parameters is regarded as a tensor product of
one dimensional modules. Similarly $q_i^{-1}$ and $e_\alpha^{-1}$ represent the dual modules and 
a sum of monomials in the equivariant parameters represents a direct sum of one dimensional modules.

We can classify the fixed points by generalizing the argument in \cite{Nak}. 
The outcome is that they are labeled by $N$-tuples of plane partitions 
$\vec{\pi}$ (three dimensional Young diagrams), 
which we call colored plane partition in this paper. 
To be more precise, in non abelian case $N > 1$
we have to assume that the vacuum expectation values $a_\alpha$ are distinct each other.
This means that the theory is in the Coulomb phase where the $U(N)$ gauge symmetry is 
completely broken. 
Let us take a basis $\{ w_\alpha \}$ of $W$ such that $U(1)^N$
acts by the multiplication of $e_\alpha^{-1} = e^{-i a_\alpha}$ 
on $w_\alpha$. This is possible, since we have assumed
that $a_\alpha \neq a_\beta$ for $\alpha \neq \beta$. 
Then we can show that
\beq
V = \oplus_{\alpha=1}^N V_\alpha, \qquad V_\alpha
 := {\mathbb C}[B_1, B_2, B_3] \cdot I(w_\alpha)~, \label{decomposition}
\eeq
where we allow that $V_\alpha = \{ 0 \}$ for some $\alpha$. 
Since $V = V_1 + V_2 + \cdots + V_N$ by the stability condition,
it is enough to show that $V_\alpha \cap V_\beta = \{ 0 \}$, if $\alpha \neq \beta$.
Let $v \in V_\alpha \cap V_\beta$ and $g_\lambda:= g(1, \lambda)$.
Then we can write $v= B_{i_1} \cdots B_{i_n} I(w_\alpha) = B_{j_1} \cdots B_{j_m} I(w_\beta)$
and by \eqref{fixB} with $q_i=1$ and \eqref{fixI},
we have both $g_\lambda v = B_{i_1} \cdots B_{i_n} g_\lambda I(w_\alpha) 
= e_\alpha^{-1} v$ and $g_\lambda v = B_{j_1} \cdots B_{j_m} g_\lambda I(w_\beta) = e_\beta^{-1} v$.
Hence $v=0$, since $e_\alpha \neq e_\beta$. 
By the vanishing theorem the condition \eqref{fixphi} is empty. 
To see the consequence of the remaining condition \eqref{fixB},
we consider the decomposition $V_\alpha = \oplus_{i,j,k \in {\mathbb Z}} 
V_\alpha (i-1, j-1, k-1)$, where the eigenspace of $g_q := g(q_i, 1)$ is
 \beq
 V_\alpha (i-1, j-1, k-1) = \{ v \in V_\alpha~\vert~ g_q v = q_1^{1-i}
 q_2^{1-j} q_3^{1-k} v \}~.
 \eeq
Then by the conditions of the fixed points,
it is easy to see that $I(w_\alpha) \in V_\alpha(0,0,0)$ and
that $B_1(V_\alpha(i,j,k)) \subset V_\alpha(i-1, j,k), 
B_2(V_\alpha(i,j,k)) \subset V_\alpha(i, j-1,k), B_3(V_\alpha(i,j,k)) \subset V_\alpha(i, j,k-1)$.
Furthermore, as was shown in \cite{CSS};
\begin{enumerate}
\item
$V(i,j,k) = \{ 0 \}$, if one of $i,j,k$ is non-positive.
\item
$\dim V(i,j,k) = 0,~\mathrm{or}~ 1$.
\item
$\dim V(i,j,k) \geq \dim V(i+1, j,k)$ and similar inequalities for $j$ and $k$.
\end{enumerate}
For proofs of these facts, we refer to \cite{CSS}. 
It is obvious that we can associate a plane partition $\pi_\alpha$ 
to the above decomposition data of $V_\alpha$.
Conversely, from an $N$-tuple of plane partitions $(\pi_1, \pi_2, \cdots \pi_N)$,
one can construct a homomorphism $g : T^3 \times U(1)^N 
\to GL(k, {\mathbb C})$ that solves the conditions \eqref{fixB} and 
\eqref{fixI}. Thus the fixed points of the toric action are isolated and 
they are labeled by colored plane partitions.

We can identify the plane partition $\pi$ with the set $\{ (i,j,k) \in {\mathbb N}^3~\vert~
k \leq h(i,j) \}$, where the height function $h(i,j) \in {\mathbb Z}_{\geq 0}$ satisfies
$h(i,j) \geq h(i+1, j), h(i,j) \geq h(i, j+1)$. The size of the plane partition is 
defined by the volume of the corresponding set $\vert\pi\vert := \sum_{(i,j)} h(i,j)$.
The size of the colored plane partition $\vec{\pi} = (\pi_1, \pi_2, \cdots, \pi_N)$ is
defined by $\vert\vec{\pi}\vert := \sum_{\alpha=1}^N \vert\pi_\alpha\vert$. 
By the localization theorem the partition function of our quiver matrix model is
expressed as a summation over colored plane partitions;
\beq
Z_{\mathrm{6D}}^{U(N)} (q_i, e_\alpha ; \Lambda) 
= \sum_{\vec{\pi}} \Lambda^{\vert\vec{\pi}\vert} N_{\vec{\pi}} (q_i, e_\alpha)~,
\eeq
where $\Lambda$ is the parameter of instanton expansion. As we will see shortly,
the size of the colored plane partition $\vert\vec{\pi}\vert$ is identified with the instanton
number $k$. The weight or the measure $N_{\vec{\pi}} (q_i, e_\alpha)$ at a fixed
point $\vec{\pi}$ is a rational function of the equivariant parameters $e_\alpha$ and $q_i$. 
It physically represents the quantum fluctuation around each fixed point. 
To compute it at $\vec{\pi}$ we decompose $V$ and $W$ 
as $T^3 \times U(1)^N$ module as follows;
\beq
W_{\vec{\pi}} = \sum_{\alpha =1}^N e_\alpha^{-1}~,  \qquad
V_{\vec{\pi}} = \sum_{\alpha=1}^N e_\alpha^{-1} \left( \sum_{(i,j,k) \in \pi_\alpha} q_1^{1-i}  q_2^{1-j} q_3^{1-k}\right)~. 
\eeq
The dual modules are
\beq
W^*_{\vec{\pi}} = \sum_{\alpha =1}^N e_\alpha~,  \qquad
V^*_{\vec{\pi}} = \sum_{\alpha=1}^N e_\alpha \left( \sum_{(i,j,k) \in \pi_\alpha} q_1^{i-1}  q_2^{j-1} q_3^{k-1}\right)~.
\eeq
These are direct sum decompositions of $V$ and $W$ at $\vec{\pi}$ into 
one dimensional $T^3 \times U(1)^N$ modules or the characters of $T^3 \times U(1)^N$.
Note that $\dim_{\mathbb C} W=N$ as it should be. Since $\dim_{\mathbb C} V=k$, we should have
$\vert\vec{\pi}\vert =k$. From the toric action \eqref{fixB}-\eqref{fixI}  we see 
the equivariant version of the deformation complex at $\vec{\pi}$ is
\beq
\begin{array}{ccccc}
 & & \mathrm{Hom}~(V_{\vec{\pi}}, V_{\vec{\pi}}) \otimes Q & & \\
 & & \oplus& & \mathrm{Hom}~(V_{\vec{\pi}},V_{\vec{\pi}}) \otimes \Lambda^2 Q \\
\mathrm{Hom}~(V_{\vec{\pi}},V_{\vec{\pi}}) & ~~\overset{\sigma}{\longrightarrow} &
\mathrm{Hom}~(V_{\vec{\pi}},V_{\vec{\pi}}) \otimes \Lambda^3 Q 
&  \overset{\tau}{\longrightarrow}~~& \oplus \\ 
 & & \oplus& & \mathrm{Hom}~(V_{\vec{\pi}},W_{\vec{\pi}}) \otimes \Lambda^3 Q \\
 & & \mathrm{Hom}~(W_{\vec{\pi}},V_{\vec{\pi}}) & & 
\end{array},
\eeq
where $Q = q_1 + q_2 + q_2$.  Hence, 
the character of the deformation complex is
\beqa
\chi_{{\vec{\pi}}} 
&=& V^{*} \otimes V \otimes (Q + \Lambda^3 Q) + W^{*} \otimes V
- V^{*} \otimes V \otimes (1 + \Lambda^2 Q) - W \otimes V^{*} \otimes \Lambda^3 Q \CR
&=& W^{*} \otimes V - W \otimes V^{*} (q_1 q_2 q_3)
- V \otimes V^{*} (1- q_1)(1- q_2) (1- q_3)~.
\eeqa
That is 
\beq
\chi_{\vec{\pi}} 
 = 
\sum_{\alpha, \beta =1}^N {e_\alpha \over e_\beta}
\left(
\sum_{ (i,j,k)\in \pi_\beta }  
\hskip-6pt
q_1^{1-i} q_2^{1-j} q_3^{1-k}
- 
\hskip-6pt
\sum_{  (r,s,t) \in \pi_\alpha }
\hskip-6pt
  q_1^r q_2^s q_3^t
- 
\hskip-6pt
\sum_{ {(r,s,t) \in \pi_\alpha \atop  (i,j,k)\in \pi_\beta} }  
\hskip-6pt
q_1^{r-i} q_2^{s-j} q_3^{t-k}
\prod_{\ell=1}^3(1-q_\ell)
\right).
\eeq
We first note that in the character $\chi_{{\vec{\pi}}}$
the number of the terms with positive coefficient and those with
negative coefficient coincide if we take the multiplicity into account. 
This is due to the fact that the formal dimension of the moduli space vanish 
and hence the character should vanish if we substitute $q_i = e_\alpha =1$. 
Therefore we can write the character as
\beq
\chi_{{\vec{\pi}}}(q_i, e_\alpha)  = \sum_{i=1}^m e^{w_i^{(+)}} - \sum_{i=1}^m e^{w_i^{(-)}}~,
\label{charactersum}
\eeq
where $e^{w_i^{(\pm)}}$ are monomials in $q_i^{\pm}$ and $e_\alpha^{\pm}$.
By the symmetry 
$\chi_{\vec{\pi}} (q_i, e_\alpha)
= - q_1q_2q_3 
\chi_{\vec{\pi}} (q_i^{-1}, e_\alpha^{-1})$,
we can set 
$e^{w_i^{(-)}} = q_1q_2q_3e^{-w_i^{(+)}}$.
Hence if $ e^{w_i^{(+)}}=e^{w_j^{(-)}}$ with $i\neq j$ then
$ e^{w_j^{(+)}}=e^{w_i^{(-)}}$.
But $ e^{w_i^{(+)}}\neq e^{w_i^{(-)}}$ in general, 
because if  $ e^{w_i^{(+)}}=e^{w_i^{(-)}}$
then $ e^{w_i^{(+)}}=\sqrt{q_1q_2q_3}$.
We will also show in Appendix B that
 $ e^{w_i^{(\pm)}}\neq(q_1q_2q_3)^n$ $(n\in\bZ)$. 
Then, according to the localization theorem the weight function is given by
\beq
N_{\vec{\pi}} (q_i, e_\alpha) = \prod_{i=1}^m \frac{\sinh w_i^{(-)}}{\sinh w_i^{(+)}}
~. \label{weight}
\eeq
Compared with the computation in \cite{CSS}, the weight \eqref{weight} computes
the so-called $K$ theoretic  version of the partition function. 
For the ADHM matrix model the $K$ theoretic version of Nekrasov's partition function 
corresponds to a five dimensional lift,
where the relation to topological string amplitudes becomes more transparent
\cite{Nek,IK-P1, IK-P2, EK1, EK2} .

When we impose the Calabi-Yau condition $\bq := \sqrt{q_1 q_2 q_3} =1$, the character 
reduces to 
\beq
\chi_{{\vec{\pi}}} 
= W^{*} \otimes V - W \otimes V^{*}
+ V \otimes V^{*}( q_1 + q_2 + q_3 - q_1^{-1} - q_2^{-1} - q_3^{-1} )~.
\eeq
Since $W^{*}(e_\alpha) = W(e_\alpha^{-1})$ and $V^{*}(q_i, e_\alpha)= V (q_i^{-1}, e_\alpha^{-1})$, we see that
under $q_i \to q_i^{-1}, e_\alpha \to e_\alpha^{-1}$, the sign of the character changes
$\chi_{{\vec{\pi}}} \to - \chi_{{\vec{\pi}}}$. Therefore we can put $w_i^{(-)} = - w_i^{(+)}$ and
hence 
\beq
N_{\vec{\pi}} (q_i, e_\alpha) = (-1)^m~.
\eeq
Though the integer $m$ may change, even if $N$ and $k$ are fixed, 
the parity of $m$ and $Nk$ agrees; $(-1)^m = (-1)^{Nk}$. Hence, 
the partition function is
\beq
Z_{\mathrm{CY3}}^{U(N)} (q_i, e_\alpha ; \Lambda) 
= \sum_{\vec{\pi}} \Lambda^{\vert\vec{\pi}\vert} (-1)^{N \vert\vec{\pi}\vert}
= \prod_{\alpha=1}^N \sum_{\pi_\alpha} u^{\vert\pi_\alpha\vert}
= M(u)^N~,
\eeq
where $u := (-1)^N \Lambda$ and $M(u)$ is the MacMahon function. 
This result was already obtained in \cite{CSS}. 
Note that the argument of the MacMahon function is not the equivariant parameters of the
toric action but the parameter of instanton expansion\footnote{However, according to \cite{TN} 
it is possible to regard the parameter $\widetilde{\Lambda}$ as a part of $\Omega$
background of 11 dimensional supergravity, or $M$ theory.}. 
The fact that the weight of each fixed point is $\pm 1$ reminds us of 
the topologically twisted ${\mathcal N}\!=\!4$ super Yang-Mills theory in four dimensions \cite{VW}. 


\Section{Computations in non Calabi-Yau case}

In the last section we have seen that the partition function reduces to a power of 
the MacMahon function if we impose the Calabi-Yau condition.
In particular, it is completely independent of both $q_i$ and $e_\alpha$.
This is a remarkable difference from Nekrasov's partition function $Z_{\mathrm{Nek}}$.
When we impose the self-duality condition $\epsilon_1 + \epsilon_2=0$,
$Z_{\mathrm{Nek}}$ is a function of $q:=e^{-g_s} = q_1 = q_2^{-1}$ and $a_\alpha$.
The leading term of the genus expansion by $g_s$ gives the Seiberg-Witten 
prepotential and the full expansion is identified with topological string 
amplitudes. For generic equivariant parameters $q_1$ and $q_2$,
it is expected that Nekrasov's partition 
function gives a certain refinement of topological string amplitudes \cite{HIV, AK1, AK2}.
Thus it is interesting to see what happens to our instanton partition function, 
if we do not impose the Calabi-Yau condition.

For non Calabi-Yau case the weight function $N_{\vec{\pi}}(q_i, e_\alpha)$ no longer takes a simple form
and is a rather complicated function. To obtain an idea on the structure of the partition function
we made some explicit computations for lower rank and lower instanton number and found that
the partition function is independent of $e_\alpha$. 
In the following by examining the residues we will confirm this 
up to three instanton number for general $N$. Based on these explicit computations
of several examples, we strongly believe that this property holds for higher instanton 
numbers and conjecture that the full partition function is given by
\beq
Z_{\mathrm{6D}}^{U(N)} = \exp 
\left(
\sum_{n=1}^\infty \frac{1}{n} F_N (q_1^n, q_2^n, q_3^n; \Lambda^n)
\right)~,
\eeq
where
\beq
F_N:= - \frac{\widetilde{\Lambda}(1+ \bq^2 + \bq^4 + \cdots + \bq^{2N-2})}
{(1-\widetilde{\Lambda})(1- \bq^{2N} \widetilde{\Lambda})} F_0 (q_1, q_2, q_3)~,
\eeq
and $\widetilde{\Lambda} = (-\bq)^{-N} \Lambda$.
For later convenience we have introduced
\beq
F_0(q_1, q_2, q_3):= \frac{(1-q_1 q_2)(1- q_2 q_3)(1- q_3 q_1)}{(1-q_1) (1-q_2) (1-q_3)}~.
\eeq
For $U(1)$ theory the same conjecture was already given by Nekrasov \cite{TN} and
the above proposal is a generalization to $U(N)$ theory. It may look that
there is only a little difference between $U(N)$ and $U(1)$ theories.
However, we would like to emphasize that it is a consequence of the crucial fact
that the partition function does not depend on the equivariant parameters from
the maximal torus $U(1)^N$, or the vacuum expectation values of Higgs scalars. 
If we impose the Calabi-Yau condition $\bq=1$, our conjecture implies
\beq
F_N = \frac{N u}{(1-u)^2},
\eeq
with $\widetilde\Lambda=(-1)^N \Lambda = u$. Thus we recover the result of the last section. 
In this sense the above instanton partition function suggests 
a generalization of Donaldson-Thomas theory to K\"ahler manifold.

Let us consider the following instanton expansion
\beqa
Z_{\mathrm{6D}}^{U(N)} &=& 1 + \sum_{k=1}^\infty \widetilde\Lambda^k Z_N^{(k)}(q_1, q_2, q_3)~, \\
F_N &=& 1 + \sum_{k=1}^\infty \widetilde\Lambda^k F_N^{(k)} (q_1, q_2, q_3)~. 
\eeqa
It is quite amusing that since
\beq
 - \frac{\widetilde{\Lambda}(1+ \bq^2 + \bq^4 + \cdots + \bq^{2N-2})}
{(1-\widetilde{\Lambda})(1- \bq^{2N} \widetilde{\Lambda})}
= \frac{-1}{1-\bq^2} \left( \frac{1}{1- \widetilde{\Lambda}} - \frac{1}{1-\bq^{2N}  \widetilde{\Lambda}} \right)
= - \sum_{k=1}^\infty \frac{1-\bq^{2Nk}}{1-\bq^2} \widetilde{\Lambda}^k~,
\eeq
the coefficients of the instanton expansion of $F_N$ take a very simple form;
\beq
F_N^{(k)} (q_1, q_2, q_3) = - \bq^{Nk-1}[Nk]_\bq \cdot F_0(q_1, q_2, q_3)~,
\eeq
where the $\bq$-integer is defined by 
\beq
[n]_\bq := \frac{\bq^n - \bq^{-n}}{\bq - \bq^{-1}} = \bq^{1-n} \frac{1- \bq^{2n}}{1- \bq^2}~.
\eeq
We have the $\bq$-binomial theorem (\cite{Mac}; Chap.I-2,Example 3), which is useful 
in the following computation,
\beq
\Exp{-\sum_{n>0} {(-z)^n\over n} \qint N{\bq^n} }
=
\prod_{\alpha=1}^N (1+z \bq^{N+1-2\alpha})
=
\sum_{k=0}^N z^k \qbin Nk\bq ,
\label{eq:qBinomialExp}%
\eeq
with
\beq
\qbin Nk\bq 
:= 
{\qintf N\bq \over \qintf {N-k}\bq \qintf k\bq },
\qquad
\qintf N\bq := \qint N\bq \qint {N-1}\bq \cdots \qint 1\bq.
\eeq
From this we obtain
\be
\bq^{k(N+1)}
\sum_{1\leq \alpha_i <\alpha_j \leq N} 
\prod_{i=1}^k \bq^{-2\alpha_i}
= 
\qbin Nk\bq,
\label{eq:qBinomial}%
\eeq
and
\beq
\qint N{\bq^2}
=
{\qbin N1\bq}^2 - 2\qbin N2\bq  ,
\label{eq:qintFormula2}%
\eeq
\beq
\qint N{\bq^3} 
=
{\qbin N1\bq}^3 - 3\qbin N1\bq \qbin N2\bq + 3\qbin N3\bq .  
\label{eq:qintFormula3}%
\eeq

In terms of $F_N^{(k)}$, the instanton expansion of the partition function is
\beqa
Z_N^{(1)} (q_1, q_2, q_3) &=& F_N^{(1)} (q_1, q_2, q_3)~, \CR
Z_N^{(2)} (q_1, q_2, q_3) &=& F_N^{(2)} (q_1, q_2, q_3) + \frac{1}{2} \left(F_N^{(1)} (q_1, q_2, q_3)\right)^2
 + \frac{1}{2} F_N^{(1)} (q_1^2, q_2^2, q_3^2)~,  \\
Z_N^{(3)} (q_1, q_2, q_3) &=& F_N^{(3)} (q_1, q_2, q_3) + F_N^{(2)} (q_1, q_2, q_3)F_N^{(1)} (q_1, q_2, q_3)  \CR
& &~~~ + \frac{1}{2} F_N^{(1)} (q_1^2, q_2^2, q_3^2) F_N^{(1)} (q_1, q_2, q_3)
+ \frac{1}{3} F_N^{(1)} (q_1^3, q_2^3, q_3^3) + \frac{1}{6} \left(F_N^{(1)} (q_1, q_2, q_3)\right)^3~.  \nonumber
\eeqa
In the following subsections we prove the conjecture up to three instanton number
for any $N$.

\subsection{One instanton}

The fixed points with $k=1$ are the colored plane partition 
$(\square, \bullet, \cdots, \bullet)$ and its cyclic permutations,
where $\square$ stands for the plane partition with unit volume.  
The character of the fixed point  $\vec{\pi}(\alpha)$ 
with $V^*_{\vec{\pi}(\alpha)}= e_\alpha$ is
\beq
\chi_{\vec{\pi}(\alpha)} = \sum_{\beta \neq \alpha} e_\alpha e_\beta^{-1} 
- \bq^2 \sum_{\beta \neq \alpha} e_\beta e_\alpha^{-1}
+ (q_1 + q_2 + q_3 - q_1 q_2 - q_2 q_3 - q_3 q_1)~,
\eeq
and
\beq
N_{\vec{\pi}(\alpha)}  (q_\ell, e_\lambda) = \bq^{- N}
\frac{(1- q_1 q_2)(1-q_2 q_3)(1 - q_3 q_1)}{(1-q_1)(1-q_2)(1-q_3)}
 \prod_{\beta \neq \alpha} \frac{e_\alpha - \bq^2 e_\beta}{e_\beta - e_\alpha}~.
\eeq
We can show that
\beq
\sum_{\alpha=1}^N \prod_{\beta \neq \alpha} \frac{e_\alpha - \bq^2 e_\beta}{e_\beta - e_\alpha} 
=(-1)^{N-1} (1 + \bq^2 + \bq^4 + \cdots + \bq^{2N-2})~. \label{limit1}
\eeq
In fact possible poles in the left hand side are at $e_\alpha = e_\beta$. But
we see that
\beq
\mathrm{Res}_{e_\alpha= e_\beta}N_{\vec{\pi}(\alpha)}
= - \mathrm{Res}_{e_\alpha= e_\beta}N_{\vec{\pi}(\beta)}~.
\eeq
Hence all the residues vanish and the left hand side is a constant in $e_\alpha$.
We may compute it by putting $e_\alpha = L^{-\alpha},~(1 \leq \alpha \leq N)$ and taking
the limit $L \to \infty$ to obtain \eqref{limit1}.
Thus we find that $Z_{N}^{(1)}$ does not depend on $e_\alpha$,
which physically means it is independent of $a_\alpha$,
or the relative distances of $N$ $D6$ branes. 
The partition function at one instanton is
\beq
Z_{N}^{(1)} = \sum_{\alpha=1}^N N_{\vec{\pi}(\alpha)}
=  (-\bq)^{- 1} (-1)^{N} [N]_\bq \cdot F_0(q_1, q_2, q_3)~,
\eeq
which proves the conjecture at one instanton.

\subsection{Two instanton}

Two instanton part of the partition function is computed as follows;
we have two types of configuration, whose characters are
$V_{\vec{\pi}(\alpha,i)} ^* := e_\alpha (1+ q_i), 1 \leq  \alpha \leq N, i=1,2,3$, which we call type I in the following
and $V_{\vec{\pi}(\alpha,\beta)} ^* := e_\alpha + e_\beta,1 \leq \alpha <  \beta \leq N$, which we call type II.

For type I we find
\beq
N_{\vec{\pi}(\alpha,i)} (q_\ell, e_\lambda) = \bq^{-2N} n_{\mathrm{I}}^{(i)} (q_\ell)
\prod_{\beta \neq \alpha} \frac{(e_\beta - e_\alpha \bq^2)(e_\beta - q_i e_\alpha \bq^2)}
{(e_\alpha - e_\beta)(q_i e_\alpha  - e_\beta)}~,
\eeq
where
\beq
 n_{\mathrm{I}}^{(i)}  (q_\ell) := 
\frac{(q_i - \prod_{j \neq i}q_j)\prod_{j \neq i}(1- q_i^2 q_j)}
{(1-q_i^2)\prod_{j \neq i}(q_i - q_j)} F_0(q_1, q_2,q_3)~.
\eeq
Similarly for the second type we have
\beq
N_{\vec{\pi}(\alpha,\beta)} (q_\ell, e_\lambda) = \bq^{-2N} n_{\mathrm{II}} (q_\ell)
\frac{\prod_{1 \leq i < j \leq 3} (e_\alpha - e_\beta q_i q_j)(e_\beta - e_\alpha q_i q_j)}
{\prod_{i=1}^3 (e_\alpha - e_\beta q_i)(e_\beta -  e_\alpha q_i)}
\prod_{\gamma \neq \alpha, \beta} \frac{(e_\gamma - e_\alpha \bq^2)(e_\gamma - e_\beta \bq^2)}
{( e_\alpha - e_\gamma)(e_\beta  - e_\gamma)}~,
\eeq
where
\beq
n_{\mathrm{II}}  (q_\ell) := F_0(q_1,q_2,q_3)^2~.
\eeq

Let us look at possible poles and residues there. 
There are poles at $ e_\alpha = e_\beta$ and $ q_i e_\alpha = e_\beta$.
Taking the relation $\bq^2 = q_1q_2 q_3$ into account, we see the relations 
\ba
\mathrm{Res}_{e_\alpha=e_\beta}
\left(
N_{\vec{\pi}(\alpha,i)} + N_{\vec{\pi}(\beta,i)}
\right)
&=& 0,
\qquad i=1,2,3,
\cr
\mathrm{Res}_{e_\alpha=e_\beta}
\left(
N_{\vec{\pi}(\alpha,\gamma)} + N_{\vec{\pi}(\beta,\gamma)}
\right)
&=& 0,
\qquad 1\leq\gamma\leq N,
\quad \gamma\neq\alpha,\beta,
\\
\mathrm{Res}_{q_i e_\alpha=e_\beta}
\left(
N_{\vec{\pi}(\alpha,i)} + N_{\vec{\pi}(\alpha,\beta)}
\right)
&=& 0. \nonumber
\ea
%
Therefore, the partition function does not depend on $e_\alpha$. 
By estimating the leading terms $e_\alpha = L^{-\alpha},~L \to \infty$,
we find the two instanton part of the partition function is
\beqa
Z_{N}^{(2)}
&=& \bq^{-2N} \left(\sum_{\alpha=1}^N \bq^{4\alpha-4} \sum_{i=1}^3 n_{\mathrm{I}}^{(i)} (q_\ell) 
+ \sum_{1 \leq \alpha < \beta \leq N} \bq^{2\alpha + 2\beta -4} n_{\mathrm{II}}(q_\ell)  \right) \CR
&=& \bq^{-2} \left( [N]_{\bq^2}\sum_{i=1}^3 n_{\mathrm{I}}^{(i)}  (q_\ell) 
+ \qbinom{N}{2} n_{\mathrm{II}}(q_\ell)  \right)~.
\eeqa
On the other hand the conjecture says
\beqa
Z_{N}^{(2)}
&=&  \bq^{-2}\left( - (1+\bq^2)[N]_{\bq^2}\cdot F_0 (q_1, q_2, q_3)
+ \frac{1}{2} [N]_\bq^2 \cdot F_0(q_1,q_2,q_3)^2 \right. \CR
& &~~~~\left. -  \frac{1}{2}  [N]_{\bq^2}\frac{(1+ q_1 q_2)(1+ q_2 q_3)(1+ q_3 q_1)}
{(1+ q_1)(1+ q_2)(1+ q_3)} F_0 (q_1, q_2, q_3) \right)~.
\eeqa
Using the identity \eqref{eq:qintFormula2}
we can see that the conjecture at two instanton reduces to the following identity;
\beq
[N]_{\bq^2}\cdot F_0 (q_1, q_2, q_3) \cdot
G_{U(1)} (q_1,q_2,q_3) =0~,
\eeq
where $G_{U(1)}=0$ is equivalent to the identity
\beqa
&& \frac{(q_1 - q_2 q_3)(1 - q_1^2 q_2)(1-q_1^2 q_3)}
{(1-q_1^2)(q_1-q_2)(q_1-q_3)} + \mathrm {(1,2,3)~cyclic}  \CR
&& = - (1+ \bq^2) - \frac{1}{2} \frac{(1+q_1 q_2)(1+q_2 q_3)(1+q_3 q_1)}
{(1+q_1)(1+q_2)(1+q_3)}
+ \frac{1}{2} \frac{(1-q_1 q_2)(1-q_2 q_3)(1-q_3 q_1)}
{(1-q_1)(1-q_2)(1-q_3)}~. \CR
&& \label{U1two}
\eeqa
The crucial point is that $N$ dependence is factored out and the remaining factor
$G_{U(1)}$ is universal in the sense
that it is independent of the rank $N$. That is what we have to prove
for general $N$ is the same as that for $U(1)$ case.
Actually the identity \eqref{U1two}
is necessary for proving the conjecture for $U(1)$ theory.
In this case type II configuration does not 
appear and the proof of the conjecture is easier. 
We note that the identity \eqref{U1two} is transformed into
the following form
\be
\sum_{i=1}^3 
{ p - q_i^2 \over 1 - q_i^2 }
\prod_{j(\neq i)} 
{ p q_i - q_j \over q_i - q_j  }
=
p(1+p)
+
{ 1\over 2} \prod_{\ell=1}^3 
{ p - q_\ell \over 1 - q_\ell }
+
{1\over 2}\prod_{i=1}^3 
{ p + q_\ell \over 1 + q_\ell }
~,
\label{U1two2}%
\ee
if $p= q_1q_2q_3$. Hence one can derive \eqref{U1two}
from the partial fraction decomposition
\ba
\prod_{\ell=1}^n 
{ pz - x_\ell \over z - x_\ell }
&=&
\sum_{i=1}^n 
{ p - x_i \over z - x_i }
\prod_{j(\neq i)} 
{ p x_i - x_j \over x_i - x_j  }
-\sum_{i=1}^{n-1} p^i,
\ea
with $n=3$ and $x_i=\pm z q_i$.

The fact that the proof is essentially reduced to abelian case
might be expected. We know that the result is independent 
of the vacuum expectation values of Higgs fields by looking at residues. This means 
the partition function does not depend on relative
distances of $D6$ branes and hence we can compute it
by taking the decoupling limit where $D6$ branes are
infinitely separated. In fact the leading term mentioned above
can be regarded as the result in this limit.

\subsection{Three instanton}

We have four types of configurations;

\begin{enumerate}
\item Type $A_1$~~~$V_{\vec{\pi}(\alpha,i)}^* 
= e_\alpha (1+ q_i + q_i^2), \quad 1 \leq \alpha \leq N,  \quad i=1,2,3$
\beq
N_{\vec{\pi}(\alpha,i)} (q_\ell, e_\lambda) = \bq^{-3N} n_{A_1}^{(i)} (q_\ell)
\prod_{\beta \neq \alpha} \frac{(e_\beta - e_\alpha \bq^2)(e_\beta - q_i e_\alpha \bq^2)
(e_\beta - q_i^2 e_\alpha \bq^2)}
{(e_\alpha - e_\beta)(q_i e_\alpha  - e_\beta)(q_i^2 e_\alpha  - e_\beta)}~,
\eeq
where
\beq
n_{A_1}^{(i)} (q_\ell) :=
\frac{(q_i - \prod_{j \neq i}q_j)(q_i^2 - \prod_{j \neq i}q_j)\prod_{j \neq i}(1- q_i^2 q_j)(1- q_i^3 q_j)}
{(1-q_i^2)(1-q_i^3)\prod_{j \neq i}(q_i - q_j)(q_i^2 - q_j)} F_0(q_1,q_2,q_3)~.
\eeq

\item Type $A_2$~~~$V_{\vec{\pi}(\alpha,i,j)}^* 
= e_\alpha (1+ q_i + q_j), \quad 1 \leq \alpha \leq N,  \quad 1 \leq i < j \leq 3$
\beq
N_{\vec{\pi}(\alpha,i,j)} (q_\ell, e_\lambda) = \bq^{-3N} n_{A_2}^{(i,j)} (q_\ell)
 \prod_{\beta \neq \alpha} \frac{(e_\beta - e_\alpha \bq^2)(e_\beta - q_i e_\alpha \bq^2)
(e_\beta - q_j e_\alpha \bq^2)}
{(e_\alpha - e_\beta)(q_i e_\alpha  - e_\beta)(q_j e_\alpha  - e_\beta)}~,
\eeq
where with $k \neq i,j $
\beq
n_{A_2}^{(i,j)} (q_\ell) : =
\frac{(1- q_i q_k)(1- q_j q_k)(1- q_i^2 q_j)(1- q_i q_j^2)(q_i - q_j^2 q_k)(q_j - q_i^2 q_k)}
{(1- q_i)(1- q_j)(q_i - q_k)(q_j - q_k)(q_i - q_j^2)(q_j - q_i^2)} F_0(q_1,q_2,q_3)~. 
\eeq

\item Type $B$~~~$V_{\vec{\pi}(\alpha,\beta,i)}^*
= e_\alpha (1+ q_i) + e_\beta, \quad 1 \leq \alpha \neq \beta \leq N,  \quad i = 1,2,3$
\beqa
& &N_{\vec{\pi}(\alpha,\beta,i)} (q_\ell, e_\lambda) = \bq^{-3N} n_{B}^{(i)} (q_\ell)
\prod_{\gamma \neq \alpha, \beta} \frac{(e_\gamma - e_\alpha \bq^2)(e_\gamma - e_\beta \bq^2)
(e_\gamma - q_i e_\alpha \bq^2)}
{(e_\alpha - e_\gamma)(e_\beta  - e_\gamma)(q_i e_\alpha  - e_\gamma)} \CR
& &\times \frac{(e_\alpha q_i - e_\beta \prod_{j \neq i}q_j)
(e_\beta - e_\alpha \bq^2)
(e_\beta - e_\alpha \prod_{j\neq i} q_j )  \prod_{j \neq i} (e_\beta - e_\alpha q_i^2 q_j) (e_\alpha - e_\beta q_i q_j)}
{(e_\alpha - e_\beta)(e_\alpha - e_\beta q_i)(e_\beta - e_\alpha q_i^2) 
 \prod_{j \neq i} ( e_\alpha q_i - e_\beta q_j) (e_\beta - e_\alpha q_j)}~, \CR
 &&
\eeqa
where 
\beq
n_{B}^{(i)} (q_\ell) :=
\frac{(q_i - \prod_{j \neq i}q_j)\prod_{j \neq i}(1- q_i^2 q_j)}
{(1-q_i^2)\prod_{j \neq i}(q_i - q_j)} 
F_0(q_1,q_2,q_3)^2 = n_{\mathrm{I}}^{(i)} (q_\ell) F_0(q_1,q_2,q_3)~.
\eeq

\item Type $C$~~~$V_{\vec{\pi}(\alpha,\beta, \gamma)}^*
= e_\alpha + e_\beta + e_\gamma, \quad 1 \leq \alpha < \beta < \gamma \leq N$
\beqa
N_{\vec{\pi}(\alpha,\beta, \gamma)} (q_\ell, e_\lambda) &=& \bq^{-3N} n_{C} (q_\ell)
\prod_{a,b = \alpha, \beta, \gamma}
\frac{\prod_{1 \leq i < j \leq 3} (e_a - e_b q_i q_j)}
{\prod_{i=1}^3(e_a - e_b q_i)} \CR
& &\times\prod_{\delta \neq \alpha, \beta, \gamma} \frac{(e_\delta - e_\alpha \bq^2)(e_\delta - e_\beta \bq^2)
(e_\delta - e_\gamma \bq^2)}
{(e_\alpha - e_\delta)(e_\beta  - e_\delta)(e_\gamma  - e_\delta)}~,
\eeqa
where
\beq
n_{C} (q_\ell) := F_0(q_1,q_2,q_3)^3 = n_{\mathrm{II}} (q_\ell) F_0(q_1,q_2,q_3)~.
\eeq
\end{enumerate}

As before all residues cancel out between two terms as follows;
\ba
\mathrm{Res}_{e_\alpha=e_\beta}
\left(
N_{\vec{\pi}(\alpha,i)} + N_{\vec{\pi}(\beta,i)}
\right)
&=& 0,
\qquad~~
\mathrm{Res}_{q_i e_\alpha=e_\beta}
\left(
N_{\vec{\pi}(\alpha,i)} + N_{\vec{\pi}(\beta,\alpha,i)} 
\right)
= 0,
\cr
\mathrm{Res}_{e_\alpha=e_\beta}
\left(
N_{\vec{\pi}(\alpha,i,j)} + N_{\vec{\pi}(\beta,i,j)}
\right)
&=& 0,
\qquad\,
\mathrm{Res}_{q_i e_\alpha=e_\beta}
\left(
N_{\vec{\pi}(\alpha,i,j)} + N_{\vec{\pi}(\alpha,\beta,j)}
\right)
= 0,
\nonumber
\\
\mathrm{Res}_{e_\alpha=e_\beta}
\left(
N_{\vec{\pi}(\alpha,\beta,i)} + N_{\vec{\pi}(\beta,\alpha,i)}
\right)
&=& 0,
\qquad
\mathrm{Res}_{q_i e_\alpha=e_\beta}
\left(
N_{\vec{\pi}(\alpha,\gamma,i)} + N_{\vec{\pi}(\alpha,\beta,\gamma)}
\right)
= 0, 
\ea
\be
\mathrm{Res}_{q_i^2 e_\alpha=e_\beta}
\left(
N_{\vec{\pi}(\alpha,i)} + N_{\vec{\pi}(\alpha,\beta,i)}
\right)
= 0,
\qquad
\mathrm{Res}_{q_i e_\alpha=q_j e_\beta}
\left(
N_{\vec{\pi}(\alpha,\beta,i)} + N_{\vec{\pi}(\beta,\alpha,j)}
\right)
= 0,
\nonumber
\ee
with $\gamma \neq \alpha, \beta$ and $j\neq i$.
Thus we can confirm that the partition function does not depend on $e_\alpha$ and
compute the partition function by taking the decoupling limit as before.
The three instanton part of the partition function is
\beqa
Z_{N}^{(3)}
&=& \bq^{-3N} (-1)^{N-1}
\left(
\sum_{\alpha=1}^N \bq^{6\alpha-6} \sum_{i=1}^3 n_{A_1}^{(i)} (q_\ell) 
+ \sum_{\alpha=1}^N \bq^{6\alpha-6}\sum_{(i,j)} n_{A_2}^{(i,j)} (q_\ell) \right. \CR
& &~~~ \left.
+ \sum_{1 \leq \alpha \neq \beta \leq N} \bq^{4\alpha + 2\beta -6} \sum_{i=1}^3 n_{B}^{(i)} (q_\ell)
+ \sum_{1 \leq \alpha < \beta < \gamma \leq N} \bq^{2\alpha + 2\beta + 2\gamma -6} \cdot n_{C} (q_\ell)
\right) \CR
&=&  \bq^{-3} (-1)^{N-1}
\left(  [N]_{\bq^3} \sum_{i=1}^3 n_{A_1}^{(i)} (q_\ell) 
+  [N]_{\bq^3} \sum_{(i,j)} n_{A_2}^{(i,j)} (q_\ell)  \right. \CR
& &~~~ \left. + \left( [N]_{\bq^2} [N]_\bq
- [N]_{\bq^3} \right) \sum_{i=1}^3 n_{B}^{(i)} (q_\ell) 
+\qbinom{N}{3}~n_{C} (q_\ell) \right)~. \CR
&&
\eeqa
The conjecture implies
\beqa
Z_{N}^{(3)} &=& \bq^{-3} \left(
- (1 + \bq^2 + \bq^4) [N]_{\bq^3}\cdot F_0(q_1,q_2,q_3) 
+ (1+ \bq^2)  [N]_{\bq^2} [N]_\bq \cdot F_0(q_1,q_2,q_3)^2 \right. \CR
& &~~+ \frac{1}{2}  [N]_{\bq^2} [N]_\bq
\frac{(1+q_1 q_2)(1+q_2 q_3)(1+q_3 q_1)}
{(1+ q_1)( 1+ q_2)(1+ q_3)}  F_0(q_1,q_2,q_3)^2 \CR
& &~~ - \frac{1}{3} [N]_{\bq^3}
\frac{(1+q_1 q_2 + q_1^2 q_2^2)(1+q_2 q_3 + q_2^2 q_3^2)(1+q_3 q_1 + q_3^2 q_1^2)}
{(1+ q_1 + q_1^2)( 1+ q_2 + q_2^2)(1+ q_3 + q_3^2)}  F_0(q_1,q_2,q_3)  \CR
& &~~ \left. - \frac{1}{6} [N]_\bq^3 F_0(q_1,q_2,q_3)^3 \right)~.
\eeqa

Using \eqref{eq:qintFormula3} and \eqref{U1two} which we have used at two instanton,
we see that the conjecture boils down to 
\beq
[N]_{\bq^3}\cdot F_0(q_1,q_2,q_3) \cdot H_{U(1)} (q_1,q_2,q_3)=0~,
\eeq
where $H_{U(1)} (q_1,q_2,q_3)=0$
is equivalent to the identity
\ba
&&
\hskip-12pt
\sum_{i=1}^3 \prod_{n=1}^2 
{ p - q_i^{n+1}\over 1 - q_i^{n+1}} 
\prod_{j(\neq i)} 
{ p q_i^n - q_j \over q_i^n - q_j }
+ 
\sum_{i<j\atop k\neq i,j} 
{ p q_i - q_k \over q_i - q_k }
{ p q_j - q_k \over q_j - q_k }
\prod_{n=1}^2  
{ p q_i^{n-1} - q_j^n \over q_i^{n-1} - q_j^n }
{ p q_j^{n-1} - q_i^n \over q_j^{n-1} - q_i^n }
\cr
&=&
p^2(1+p+p^2)
+
p(1+p)
f(p,q_\ell)
+
{1\over 2}
f(p^2,q_\ell^2)
+
{1\over 3}
{ f(p^3,q_\ell^3) \over f(p,q_\ell) }
+
{1\over 3!}
f(p,q_\ell)^2
~, 
\label{U1three}%
\ea
with $f(p,q_\ell):=\prod_{\ell=1}^3{( p - q_\ell )/( 1 - q_\ell )}$,
if $p = q_1q_2q_3$.
Again we can factor out $N$ dependence completely and what we have to show
is the identity \eqref{U1three}, which is required for proving 
the conjecture for $U(1)$ theory. Note that in $U(1)$ case
the colored plane partitions of type $B$ and $C$ do not appear. 
We can check the identity \eqref{U1three} by direct computation
based on the partial fraction decomposition.

In summary, computations of instanton number two and three show that
basic ingredients for the validity of the conjecture are
identities for $\bq$-integers such as \eqref{eq:qintFormula2} and \eqref{eq:qintFormula3}
and the combinatorial identity for $U(1)$ theory like \eqref{U1two} and \eqref{U1three}.
We believe we will see similar structure for higher instanton numbers.
In fact \eqref{eq:qintFormula2} and \eqref{eq:qintFormula3} are the first two identities 
which are derived form the $\bq$-binomial theorem
\eqref{eq:qBinomialExp}.
On the other hand at the moment we cannot see any underlying reason for the identities \eqref{U1two} and
\eqref{U1three}, though we can check them by considering the partial fraction decomposition.
Since they are the equalities for $U(1)$ theory, it is tempting to expect
that they are related to the geometry or combinatorics of 
the Hilbert scheme $\mathrm{Hilb}^n ({\mathbb C}^3)$ of points in ${\mathbb C}^3$.

%

\section*{Acknowledgments}


We would like to thank H.~Fuji, M.~Hamanaka, M.~Manabe, S.~Moriyama, H.~Ochiai
and M.~Shimizu for discussions. We also thank K.~Ohta for his
inspiring talk at Yukawa Institute, Kyoto in March 2009. 
This work is partially supported by the Grant-in-Aid for Nagoya
University Global COE Program, "Quest for Fundamental Principles in the
Universe: from Particles to the Solar System and the Cosmos", from the Ministry
of Education, Culture, Sports, Science and Technology of Japan.
The present work is also supported in part by Daiko Foundation. 
The work of H.K.  is supported in part by Grant-in-Aid for Scientific Research
[\#19654007] from the Japan Ministry of Education, Culture, Sports, Science and Technology

\Section*{Appendix A : ADHM matrix model and Nekrasov's partition function}
\renewcommand{\theequation}{A.\arabic{equation}}\setcounter{equation}{0}
\renewcommand{\thesubsection}{A.\arabic{subsection}}\setcounter{subsection}{0}

In this appendix we review how we can derive Nekrasov's instanton partition function
as an equivariant index of the matrix quantum mechanics of the ADHM equations.
Let us consider two vector spaces $V$ and $W$ with complex dimensions,
$\dim_{\mathbb C} V=k$ and $\dim_{\mathbb C} W= N$.
In the language of $D$ brane system we have $k$ $D0$ branes bound to $N$ $D4$ branes.
As an effective theory on $D4$ branes we have $U(N)$ gauge theory and 
$k$ $D0$ branes describe the gas of point-like $k$ instantons. 
In the $D$ brane picture the ADHM construction is a dual description where
we consider an effective $0+1$ dimensional theory on $D0$ brane
\cite{WitADHM1, WitADHM2, Doug1, Doug2}.
We have $B_{1,2} \in \mathrm{Hom}~(V,V)$ from \lq\lq 0-0\rq\rq\ string.  
From \lq\lq 0-4\rq\rq\ and \lq\lq 4-0\rq\rq\ string
we have $I \in \mathrm{Hom}~(W,V)$ and $J \in \mathrm{Hom}~(V,W)$.
The ADHM equations for these ADHM data are
\beqa
{\cal E}_{\mathbb C} &:=& [B_1, B_2] +IJ = 0~, \\
{\cal E}_{\mathbb R}(\zeta) &:=&[B_1, B_1^\dagger] + [B_2, B_2^\dagger] + I I^\dagger - J^\dagger J - \zeta =0~.
\eeqa
When we construct the moduli space of instantons as the hyperK\"ahler quotient, 
they play the role of hyperK\"ahler moment maps.
Namely the moduli space can be identified with 
\beq
{\cal M}_{\mathrm{ADHM}} := \{ (B_1, B_2, I,J) \vert~
{\cal E}_{\mathbb C}=0,  {\cal E}_{\mathbb R}(\zeta)=0 \}/ U(k)~.
\eeq
The formal dimension of ${\cal M}_{\mathrm{ADHM}}$ is computed as follows;
we impose $2k^2 + k^2$ (real) constraints on 
$4k^2 + 4NK$ (real) degrees of freedom from the matrices $(B_1, B_2, I, J)$.
Since the gauge group $U(k)$ reduces further $k^2$ degrees of freedom,
we find the moduli space has $4NK$ dimensions, or $\dim_{\mathbb C} 
{\cal M}_{\mathrm{ADHM}} =2Nk$, which agrees to the dimensions of
the moduli space of ASD instanton of $U(N)$ theory with instanton number $k$.
It is known that the moduli space is isomorphic to the following affine algebro-geometric
quotient \cite{Nak, NY1};
\beq
\widetilde{\cal M}_{\mathrm{ADHM}} := \{ (B_1, B_2, I,J) \vert~
{\cal E}_{\mathbb C}=0  \}/\!/ GL(k, {\mathbb C})~. \label{ADHMquotient}
\eeq
In \eqref{ADHMquotient} instead of the $D$ term condition we impose the algebraic 
stability condition that there is no proper subspace $S$ of $V$ which satisfies
$B_1 S \subset S, B_2 S \subset S$ and $\mathrm{Im}~(I) \subset S$.

We consider the toric action $(z_1, z_2) \to (e^{i\epsilon_1} z_1, e^{i\epsilon_2} z_2)$
of $T^2$ on ${\mathbb C}^2$.
The ADHM data transform $(B_1, B_2, I, J) \to ( q_1\cdot B_1, q_2 \cdot B_2, I, (q_1q_2)\cdot J)$
where $q_i := e^{i\epsilon_i}$. The fixed points are isolated and classified by $N$-tuples of
partitions $\vec{\lambda}$ \cite{Nak}. 
The equivariant deformation complex at a fixed point $\vec{\lambda}$ is \cite{Nak, NY1, FP, BFMT, LMN};
\beq
\begin{array}{ccccc}
 & & \mathrm{Hom}~(V_{\vec{\lambda}}, V_{\vec{\lambda}}) \otimes Q & & \\
 & & \oplus & & \\
\mathrm{Hom}~(V_{\vec{\lambda}},V_{\vec{\lambda}}) &
~~\overset{\sigma}{\longrightarrow}  & \mathrm{Hom}~(W_{\vec{\lambda}},V_{\vec{\lambda}}) 
& \overset{\tau}{\longrightarrow}~~&   \mathrm{Hom}~(V_{\vec{\lambda}},V_{\vec{\lambda}}) \otimes \Lambda^2 Q \\ 
 & & \oplus & & \\
 & & \mathrm{Hom}~(V_{\vec{\lambda}},W_{\vec{\lambda}}) \otimes \Lambda^2 Q & & 
\end{array},
\eeq
where $Q = T_1^{-1}  + T_2^{-1}$ and $T_i$ is one dimensional module
on which $T^2$ acts as the multiplication of $e^{i\epsilon_i}$.
Hence the equivariant index is
\beqa
\chi &=& (V^* \otimes V) Q + W^* \otimes V + V^* \otimes W  \otimes \Lambda^2 Q  - (V^* \otimes V) (1+  \Lambda^2 Q) \CR
&=&  W^* \otimes V + V^* \otimes W  (T_1 T_2)^{-1} -  V^* \otimes V (1- T_1^{-1}) (1- T_2^{-1})~. \label{4Dcharacter}
\eeqa
We have $2NK$ positive terms in this index which are regarded as the weights (eigenvalues) of 
the toric action at the fixed points\footnote{In the character 
\eqref{4Dcharacter} all the term with negative coefficient are canceled and 
there are $2Nk$ remaining terms.}. 
Each weight is a monomial in the equivariant
parameters $q_i^\pm = e^{ \pm i \epsilon_i}$ from $T^2$ and $e_\alpha^\pm= e^{\pm i a_\alpha}$.
Hence from a character of the form $\chi = \sum_{i=1}^{2Nk} \exp(w_i)$,
we obtain the following contribution to the instanton partition function;
\beq
z(\vec{\lambda}) = \prod_{i=1}^{2Nk} ( 1- \exp(w_i))^{-1}~,
\eeq
where we consider the $K$ theoretic version of the partition function,
which corresponds to the index of the Dolbeault operator $\bar\partial$
or the Todd class. 
By localization theorem the partition function is
computed by summing up all the contributions at each fixed point, or
the colored partition $\vec{\lambda}$;
\beq
Z_{\mathrm{Nek}} (e_\alpha, q_i ; \Lambda) 
= \sum_{\vec{\lambda}} \left(\frac{\Lambda} {\sqrt{q_1 q_2}} \right)^{N \vert \vec{\lambda} \vert}  
\frac{1}{\prod_{\alpha, \beta =1}^N N_{\alpha, \beta} (e_\alpha, q_i)}~,
\eeq
where 
\beq
N_{\alpha, \beta} (e_\alpha, q_i) = \prod_{s \in \lambda_\alpha}
\left( 1- q_1^{-\ell_{\lambda_\beta} (s) -1} q_2^{a_{\lambda_\alpha}(s)} e_\alpha e_\beta^{-1} \right)
\prod_{t \in \lambda_\beta}
\left( 1- q_1^{\ell_{\lambda_\alpha} (t)} q_2^{-a_{\lambda_\beta}(t) -1} e_\alpha e_\beta^{-1}\right)~.
\eeq
Note that we have renormalized the parameter $\Lambda$ of instanton expansion
by $\sqrt{q_1 q_2}$ as we made for the topological partition function in this paper.

\Section*{Appendix B : Well-definedness of $N_{\vec{\pi}} (q_i, e_\alpha)$}
\renewcommand{\theequation}{B.\arabic{equation}}\setcounter{equation}{0}
\renewcommand{\thesubsection}{B.\arabic{subsection}}\setcounter{subsection}{0}

\newcommand{\nst}[2]{n(#1;#2)}
\newcommand{\nsi}[2]{n_{#2}(#1)}

To define the weight function $N_{\vec{\pi}} (q_i, e_\alpha)$, 
$\{e^{w_i^{(+)}}\}$, which is defined by \eqref{charactersum},
 should not contain $1$. We prove it here.


A plane partition $\pi$ is define as a finite set of positive integers,
$\pi=\{(i,j,k)\} \subset \bN^3$, such that
if $(i,j,k)\in\pi$ then $(i',j',k')\in\pi$ 
($1\leq i' \leq i$, $1\leq j' \leq j$, $1\leq k' \leq k$). 
Given any plane partition $\pi$,
let 
\be
\nst st
:= 
\#\Setv
{(i,j,k)\in\pi}
{(i',j',k'):=(i-s_1-t_1,j-s_2-t_2,k-s_3-t_3)\in\pi},
\ee
with $s=(s_1,s_2,s_3)$, $t=(t_1,t_2,t_3)$ and
\ba
\nsi s0 
&:=&
\nst s{0,0,0},
\qquad
\nsi s{-1}  
:= 
\#\left\{ 
(s_1+1,s_2+1,s_3+1)\in\pi
\right\},
\cr
\nsi s1  
&:=& 
\nst s{1,0,0}
+\nst s{0,1,0}
+\nst s{0,0,1},
\cr
\nsi s2 
&:=& 
 \nst s{0,1,1}
+\nst s{1,0,1}
+\nst s{1,1,0},
\cr
\nsi s3 
&:=& 
\nst s{1,1,1}.
\ea
Note that
\be
\nsi{0,0,0}0 
=
|\pi |,
\qquad
\nsi{0,0,0}{-1}  
= 
\left\{ 
\begin{array}{ll}
0,\qquad &\pi = \emptyset \\
1,\qquad &\pi\neq \emptyset
\end{array}
\right.
.
\ee
First we have
\\
{\bf Lemma.}~{\it
If
$(s_1,s_2,s_3)\in\bZ_{\geq 0}^3$
then
$\sum_{\ell=-1}^3 (-1)^\ell \nsi s{\ell} = 0$.
}

\proof
For $\pi=\emptyset$, since $\nsi s{\ell} = 0$,
the lemma holds.
Assuming the lemma to hold for $\pi$,
we will prove it for the plane partition $\pi' = \pi \cup \{(i,j,k)\}$.
The differences between $\nsi s{\ell}$'s of $\pi$ and those of $\pi'$, 
$
({\it \Delta} \nsi s{-1} ,{\it \Delta} \nsi s0,
{\it \Delta} \nsi s1,{\it \Delta} \nsi s2, {\it \Delta} \nsi s3)
$,
are
\be
\begin{array}{ll}
(0,1,3,3,1), \qquad &  i-s_1,j-s_2,k-s_3 > 1,
\cr
(0,1,2,1,0), \qquad & \{ i-s_1,j-s_2,k-s_3\} = \{1, \alpha ,\beta \},
\cr
(0,1,1,0,0), \qquad & \{ i-s_1,j-s_2,k-s_3\} = \{1, 1, \alpha \},
\cr
(1,1,0,0,0), \qquad & i-s_1=j-s_2=k-s_3=1,
\cr
(0,0,0,0,0), \qquad & i-s_1 \quad{\rm or}\quad j-s_2 \quad{\rm or}\quad k-s_3 < 1,
\end{array}
\ee
with
$\alpha ,\beta > 1$.
Thus it holds for $\pi'$.
\qed

For $e^{w_i^{(\pm)}}$ introduced in \eqref{charactersum} and \eqref{weight}, we have
\\ 
{\bf Proposition.}~{\it
$e^{w_i^{(\pm)}} \neq q_1^{n_1}q_2^{n_2}q_3^{n_3}$ 
with
$(n_1,n_2,n_3)\in\bZ_{\leq 0}^3$ or $\in\bN^3$.
}

\proof
It suffices to show it when $N=1$, i.e., for
\ba
\chi_\pi(q_i,e_1)
& = &
\sum_{ (i,j,k)\in \pi }  q_1^{1-i} q_2^{1-j} q_3^{1-k}
-
\sum_{  (i',j',k') \in \pi }  q_1^{i'} q_2^{j'} q_3^{k'}
\cr
&-&
\sum_{ (i,j,k), (i',j',k') \in \pi }  q_1^{i'-i} q_2^{j'-j} q_3^{k'-k}
\left(
1-\sum_{\ell=1}^3 q_\ell + \sum_{\ell=1}^3{q_1q_2q_3\over q_\ell} 
- q_1q_2q_3
\right).
\label{eq:ChiN1}%
\ea
Each monomial $q_1^a q_2^b q_3^c$  
in the 1st, 3rd, 4th, 5th and 6th terms of \eqref{eq:ChiN1}
becomes $q_1^{-s_1}q_2^{-s_2}q_3^{-s_3}$ ($s_i \in\bZ_{\geq 0}$) 
if and only if
$(i,j,k)=(s_1+1,s_2+1,s_3+1)$,
$(i,j,k)-(i',j',k')=(s_1,s_2,s_3)$,
\ba
(i,j,k)-(i',j',k')-(s_1,s_2,s_3)
&=&(1,0,0)\quad{\rm or}\quad(0,1,0)\quad{\rm or}\quad(0,0,1),
\cr
(i,j,k)-(i',j',k')-(s_1,s_2,s_3)
&=&(0,1,1)\quad{\rm or}\quad(1,0,1)\quad{\rm or}\quad(1,1,0),
\cr
(i,j,k)-(i',j',k')-(s_1,s_2,s_3)&=&(1,1,1),
\ea
respectively.
But the number of them are
$\nsi s{-1}$, $\nsi s0$, $\nsi s1$, $\nsi s2$ and $\nsi s3$, 
respectively, whose alternating summation vanishes.
Thus 
$e^{w_i^{(\pm)}} \neq q_1^{n_1}q_2^{n_2}q_3^{n_3}$ 
with
$(n_1,n_2,n_3)\in\bZ_{\leq 0}^3$.
The symmetry 
$\chi_{\vec{\pi}} (q_i, e_\alpha)
= - q_1q_2q_3 
\chi_{\vec{\pi}} (q_i^{-1}, e_\alpha^{-1})$
guarantees that 
$e^{w_i^{(\pm)}} \neq q_1^{n_1}q_2^{n_2}q_3^{n_3}$ 
with
$(n_1,n_2,n_3)\in\bN^3$.
\qed

Therefore, $N_{\vec{\pi}} (q_i, e_\alpha)$ is well-defined.


\end{document}